\documentclass[11pt]{article}
\usepackage[T1]{fontenc}
\usepackage[utf8]{inputenc}
\usepackage[margin=1in]{geometry}
\usepackage{amsmath,amssymb,amsthm}
\usepackage{graphicx,placeins,flafter}
\usepackage[hidelinks]{hyperref}
\usepackage{microtype,lmodern,xcolor}
\newtheorem{theorem}{Theorem}[section]
\newtheorem{proposition}[theorem]{Proposition}
\newtheorem{lemma}[theorem]{Lemma}
\theoremstyle{definition}
\newtheorem{definition}[theorem]{Definition}
\newif\ifmarked
\markedfalse

\title{Revisiting Unicity Distance: A Channel Transmission Perspective and Its Limits}
\author{Fangyuan Lin\\[0.3em]\small Department of Statistics, Columbia University}
\date{}
\begin{document}
\maketitle
\begin{abstract}
This paper revisits the classical notion of unicity distance for substitution ciphers from an information-theoretic perspective. We frame simple-substitution encryption as a noisy transmission channel, in which the plaintext is the message, the unknown key is the noise, and the attacker is the receiver. Using typical sets to motivate the notion of meaningful plaintexts, we apply the channel coding theorem to blocks of plaintext encrypted with independent fresh keys. We show that a block length exceeding $H(K)/D$ permits reliable communication at the language's entropy rate using suitably chosen codebooks across many blocks. This provides a channel transmission interpretation of the classical unicity-distance formula. We then test how well this formula describes plaintext recovery when the message is a passage from six known novels. More than $99\%$ of the passages are uniquely identifiable at 18 letters, well before the classical estimate of about 28. N-gram language models, based on the frequencies of short letter sequences, predict the recovery probabilities much more closely. Hellman (1977) warned that the classical estimate can overstate security, but not by how much. Here it is off by about ten letters.
\end{abstract}

\section{Introduction}
\label{sec:Introduction}
Shannon's unicity distance is the approximate amount of
ciphertext at which a unique meaningful decipherment is
expected \cite{shannon1949}. In cryptography, encryption is a process that transforms a message, called plaintext, into a ciphertext. Encryption is represented as a function that maps the plaintext $X$ and a key $K$ to the ciphertext $Y$: 
\[
Y:=\mathrm E_K(X).
\]
We consider symmetric encryption, for which decryption with the same key recovers the original plaintext:
\[
X=\mathrm D_K(Y).
\]
The attacker observes the ciphertext and knows the encryption scheme, but not the key.

This paper studies the simple substitution cipher on the $26$ English letters. Its key is a permutation of the alphabet, and encryption replaces every letter of the plaintext by its image under that permutation. Intuitively, a short ciphertext often admits several readings, while a longer one can distinguish its plaintext. One classical attack is frequency analysis: in English the letter \texttt{E} is the most common, so the most frequent symbol in a long ciphertext probably stands for \texttt{E}. With only a few letters, such counts are unreliable and decipherment may remain ambiguous. For example the ciphertext \texttt{XYZ} could represent the meaningful plaintexts \texttt{CAT} or \texttt{DOG}, and the ciphertext alone would not distinguish the two. On the other hand, additional letters can leave only one sensible plaintext, and we ask how the probability of this event depends on ciphertext length.

We revisit unicity distance \cite{shannon1949} by viewing encryption as channel transmission. In this interpretation, an entire plaintext block is the input to a channel, the unknown random key supplies its noise, and the attacker attempts to decode its output. Independently encrypted blocks form uses of a discrete memoryless channel (DMC) \cite[p.~193]{cover2006}; letters within one block share a key and are not independent channel uses. Shannon made this connection in his description of secrecy systems as noisy communication systems \cite[pp.~685--686]{shannon1949}. Throughout, the attacker's goal is to recover the plaintext and as we will see, this does not always require recovering the entire key.

In our formulation, the informal notion of a meaningful message is made precise using the \emph{typical set}. Hellman \cite{hellman1977} noted that the notion ``meaningful''
can almost be replaced by ``typical'' in an ergodic source
model and we discuss in Section~\ref{sec:proof} the introduction of typicality in detail. The channel coding theorem \cite{cover2006} guarantees reliable communication using suitable codes across sufficiently many channel uses; we show that $ND>H(K)$ makes the rate $NR$ achievable by suitable codes over independently encrypted blocks (Section~\ref{sec:capacity}). In Section~\ref{sec:experiment}, we compute the probability of unique decipherment for passages from six English novels and compare it with predictions from N-gram models using public letter-transition tables.

I began this project as an undergraduate, hoping that information and communication theory would give a single answer to the decipherment question. Working through the experiment changed my view. The classical formula does not describe the observed transition from multiple possible plaintexts to unique decipherment. Simple models of English letter sequences do much better. In this paper, we use the channel coding theorem to recover the classical
expression as a sufficient condition for the substitution cipher to become reliable communication,
then examine how well it describes plaintext recovery
in a fixed collection of texts (Section~\ref{sec:experiment}).

Previous work has studied spurious decryptions, uncertainty about the
key, and the performance of decipherment algorithms
\cite{hellman1977,blom1979,hasinoff2003,norvig2009,raviknight2008}.
Our experiment asks whether the plaintext is uniquely determined,
regardless of the decoding algorithm. We treat passages from six known
novels as the complete set of possible plaintexts. By checking every
candidate, we obtain the exact probability of unique decipherment
within this collection at each ciphertext length. This gives a
benchmark against which to compare the classical unicity-distance
estimate and predictions from N-gram language models. 

\section{Background: Entropy and Redundancy of Language}
Let $X$ be a discrete random variable taking values in a finite set
$\mathcal X$, called the \emph{alphabet} of $X$. The entropy of $X$,
denoted $H(X)$, measures the uncertainty or information content of $X$
\cite{shannon1951}:
\[
H(X) = -\sum_{x\in \mathcal{X}} \Pr(X=x) \log_2 \Pr(X=x).
\]
An \emph{information source} is a sequence of random variables
$\{X_1,X_2,\dots\}$ taking values in a common alphabet $\mathcal X$. For plaintext, $\mathcal X$ is the set of letters, of size $G\ge2$
($G=26$ for English). By the independence and uniform bounds \cite[Theorems~2.6.4 and 2.6.6]{cover2006},
\[
H(X^N)\le\sum_{i=1}^NH(X_i)\le N\log_2G,
\]
with equality throughout if and only if the letters are independent and uniformly distributed. We write $R_0=\log_2G$ for this largest possible entropy per letter, called the \emph{absolute rate}.

For a stationary information source, the limits
\[
R=\lim_{N\to\infty}\frac1NH(X^N)
=\lim_{N\to\infty}H(X_N\mid X_{N-1},\dots,X_1)
\]
exist and are equal \cite[Theorems~4.2.1 and 4.2.2]{cover2006}. We call $R$ the \emph{entropy rate} and $D=R_0-R$ the \emph{redundancy} per letter. Linguistic conventions, such as ``q'' being usually followed by ``u,'' make text predictable and create redundancy. We use the frequently quoted value $R=1.5$ bits per letter for English \cite{shannon1951,hellman1977}, which gives $D\approx3.20$ bits per letter; it serves as a model parameter.

\subsection{Classical Result: Unicity Distance}

A \textit{spurious key} is an incorrect key that also yields a meaningful decryption. For example, if the plaintext is \textit{cat} and it is encrypted into ciphertext $c$, then the key that deciphers $c$ into \textit{dog} would be a spurious key, since \textit{dog} is also a valid English word but it is not the actual plaintext. Intuitively, if the plaintext is ``too short," then several meaningful decryptions may remain, and the question we ask is how short is ``too short." To address this issue, we study the concept of unicity distance, which was first introduced by Shannon \cite{shannon1949}. The classical approximate \textit{unicity distance} is the crossover length at which the expected number of spurious keys falls to order one. The random-cipher calculation underlying the classical approximation uses the following assumptions \cite{shannon1949,hellman1977}:

\begin{enumerate}
    \item There are $G^N=2^{NR_0}$ strings of length $N$ over the alphabet $\mathcal X$ of size $|\mathcal X| = G$.
\item There are $M$ allowed plaintexts, sampled uniformly. At this fixed length, write $R=(\log_2 M)/N$ and $D=R_0-R$, where $D$ is called the redundancy of the language.
    \item There are $|\mathcal K|$ keys, with key random variable $K$ a priori distributed uniformly and independent of the plaintext variable. Each key independently specifies a uniformly random bijection of the entire space of length-$N$ strings.
\end{enumerate}

By \cite{hellman1977}, under these assumptions, the expected number $\widehat n_k$ of spurious keys is
\begin{equation}\label{eq:hellman}
\widehat n_k=(2^{H(K)}-1)\frac{M}{G^N}
=(2^{H(K)}-1)2^{-ND}\approx2^{H(K)-ND}.
\end{equation}
The crossover where this expectation is approximately one gives the classical approximate formula
\begin{equation}\label{eq:classical}
U\approx\frac{H(K)}{D}.
\end{equation}
It does not assert that the expected number of spurious keys becomes exactly zero. Beauchemin and Brassard \cite{beauchemin1988} extended Hellman's formula to nonuniform plaintext and key distributions.

For a uniform simple-substitution key, $H(K)=\log_2(26!)\approx88.38$ bits. Inserting $R=1.5$ into \eqref{eq:classical} gives approximately $27.62\approx 28$ letters. We now recover this expression through a communication-theoretic approach.

\section{Encryption as a Communication Channel}
\label{sec:proof}
We use the language of channel coding. A \emph{codebook} is a list $\mathcal C=(c_1,\ldots,c_M)$ of possible plaintexts of length $N$, known to both the sender and the attacker; its entries are called \emph{codewords}. The sender chooses a message index $J$ uniformly from $\{1,\ldots,M\}$ and encrypts the codeword $c_J$. The attacker observes the ciphertext and tries to identify the transmitted codeword. The \emph{rate} of the codebook is $(\log_2M)/N$ bits per letter. Thus meaningful plaintexts are codewords, encryption is transmission, and decipherment is decoding.

\subsection{Typical Sets and Meaningful Plaintexts}\label{sec:typical}
We informally call a plaintext meaningful when it is plausible under a statistical model of language. Let $\{X_1,X_2,\dots\}$ be a stationary ergodic information source with alphabet $\mathcal X$ and entropy rate $R$.

For a stationary information source $\{X_1,X_2,\dots\}$ on a finite alphabet $\mathcal X$, the limits
\[
R=\lim_{N\to\infty}\frac1N H(X_1,\dots,X_N)
 =\lim_{N\to\infty}H(X_N\mid X_{N-1},\dots,X_1)
\]
exist and are equal \cite[Theorems~4.2.1--4.2.2]{cover2006}. We call $R$ the
\emph{entropy rate} of the source and $D=\log_2G-R$ its redundancy per letter,
consistent with the notation of Section~2. Since $\frac1N H(X_1,\dots,X_N)$
is nonincreasing in $N$ and converges to $R$, the per-letter entropy of a finite
block may exceed $R$; the count $2^{NR}$ below is therefore an asymptotic
description.

\begin{definition}[Typical set]\label{def:typical}
For $\varepsilon>0$, the $\varepsilon$-typical set of length $N$ is
\[
\mathcal T_N(\varepsilon)=\left\{x^N\in\mathcal X^N:
\left|-\frac1N\log_2\Pr\{X^N=x^N\}-R\right|\le\varepsilon\right\}.
\]
\end{definition}

By the asymptotic equipartition property for stationary ergodic sources (the Shannon--McMillan--Breiman theorem) \cite{cover2006}, $\Pr\{X^N\in\mathcal T_N(\varepsilon)\}\to1$ as $N\to\infty$, and for large $N$
\[
(1-\varepsilon)2^{N(R-\varepsilon)}\le|\mathcal T_N(\varepsilon)|\le2^{N(R+\varepsilon)}.
\]
Thus about $2^{NR}$ texts carry nearly all the probability, which is a small fraction of the $2^{NR_0}$ strings when $R<R_0$. Each typical text has probability between $2^{-N(R+\varepsilon)}$ and $2^{-N(R-\varepsilon)}$, that is, about $2^{-NR}$ to first order in the exponent. Following \cite{hellman1977}, we use typicality to motivate a \emph{codebook} of about $2^{NR}$ meaningful plaintexts, used with equal probability.

Although stationarity and ergodicity are idealizations of natural language, Shannon's prediction experiments provided an influential estimate of the entropy of English: between about $0.6$ and $1.3$ bits per letter when the preceding $100$ letters are known \cite{shannon1951}. The value $R=1.5$ used in this paper is the conventional, slightly conservative figure corresponding to the redundancy $D\approx3.2$ bits per letter quoted by \cite{hellman1977}.

\subsection{The Substitution Channel}\label{sec:channel}
Having described the messages, we now turn to the channel that carries them. In communication theory, a \emph{channel} is a random mechanism that turns an input into an output. It is specified by an input alphabet $\mathcal A$, an output alphabet $\mathcal B$, and a \emph{transition law} $W(y\mid x)$, the probability that the output is $y\in\mathcal B$ when the input is $x\in\mathcal A$; for each $x$, $W(\cdot\mid x)$ is a probability distribution on $\mathcal B$. A noisy telephone line is the standard example: the speaker's words are the input, what the listener hears is the output, and noise on the line makes the output random. The receiver sees only the output and must infer the input.

The key observation of this paper is that encryption fits this description once we take the attacker's point of view. Shannon made the same observation: ``From the point of view of the cryptanalyst, a secrecy system is almost identical with a noisy communication system,'' he wrote on page~685 of \cite{shannon1949}. The rest of this section makes this precise for substitution cipher. Let $\mathcal X$ have $G\ge2$ letters. Choose $\pi_K$ uniformly among the $G!$ permutations of $\mathcal X$, independently of $X^N$, and encrypt letter by letter: $Y_i=\pi_K(X_i)$ for $i=1,\ldots,N$. For the intended recipient, who knows $K$, this map is deterministic and can be inverted. The attacker does not know $K$, so to the attacker the key acts as random noise. The input is the whole plaintext $X^N\in\mathcal X^N$, the output is the whole ciphertext $Y^N\in\mathcal X^N$, and averaging over the unknown key gives the transition law of the channel
\begin{equation}\label{eq:channel-average}
W_N(y^N\mid x^N)=\frac1{G!}\sum_k
\mathbf1\{y^N=\pi_k(x^N)\},
\end{equation}
the probability that the plaintext $x^N$ is encrypted to $y^N$. We call $W_N$ the \emph{substitution channel}. Because one key encrypts every letter, the letters of a ciphertext are not independent uses of a letter-by-letter channel. Instead, the input and output symbols of $W_N$ are whole length-$N$ blocks, and one use of $W_N$ encrypts an entire plaintext.

What does this channel preserve? A substitution changes the names of letters but never which positions hold the same letter. Let $\operatorname{pat}(x^N)$ record which positions contain the same letter, labeling distinct letters in order of first appearance. For example, \texttt{EYE} and \texttt{DID} have pattern $(1,2,1)$, whereas \texttt{TOO} has pattern $(1,2,2)$. A substitution channel maps one plaintext to a ciphertext exactly when their patterns agree; we then say they are \emph{compatible}. If $x^N$ contains $d(x^N)$ distinct letters, the keys mapping $x^N$ to a compatible $y^N$ are those that agree on these letters, and there are $(G-d(x^N))!$ of them. Therefore
\begin{equation}\label{eq:channel}
W_N(y^N\mid x^N)=
\begin{cases}
(G-d(x^N))!/G!,&\operatorname{pat}(y^N)=\operatorname{pat}(x^N),\\
0,&\text{otherwise}.
\end{cases}
\end{equation}
All plaintexts compatible with a ciphertext have the same likelihood. A decoder can therefore identify the compatible codewords but cannot distinguish their message indices by likelihood. We call plaintext decipherment \emph{unique} when all compatible entries represent the same plaintext value, and \emph{ambiguous} otherwise. For a list of distinct plaintexts this means that exactly one entry is compatible with the ciphertext.

To decide when the list shrinks to one entry, we need a measure of how much evidence a ciphertext provides. For an input law $Q$ on $\mathcal X^N$, the output law and the \emph{information density} are
\[
Q_Y(y)=\sum_xQ(x)W_N(y\mid x),\qquad
\imath_Q(x;y)=\log_2\frac{W_N(y\mid x)}{Q_Y(y)}.
\]
The information density measures how much more likely the observed ciphertext is under the transmitted plaintext than under the output law. An information density of $b$ bits means that the ciphertext is $2^b$ times more probable given the true plaintext than it is in general. The average of the information density is the mutual information $I_Q(X^N;Y^N)$. The mutual information
\[
I(X;Y)=\sum_{x,y}\Pr\{X=x,Y=y\}\log_2\frac{\Pr\{X=x,Y=y\}}{\Pr\{X=x\}\Pr\{Y=y\}}=H(X)-H(X\mid Y)
\]
is the reduction in uncertainty about $X$ by observing $Y$ \cite{cover2006}. It is symmetric, $I(X;Y)=I(Y;X)=H(Y)-H(Y\mid X)$, and it is zero exactly when $X$ and $Y$ are independent. When the distribution of $X$ is a chosen input law $Q$, we write $H_Q$ and $I_Q$ to make the dependence explicit.

The following proposition makes precise the observation that the substitution channel transmits patterns and nothing else.

\begin{proposition}[Information conveyed by substitution]\label{prop:information}
For any input law $Q$ independent of the key, let $S=\operatorname{pat}(X^N)$ have distribution $Q_S$. Then $\imath_Q(X^N;Y^N)=-\log_2Q_S(S)$ almost surely, and
\begin{equation}\label{eq:information}
I_Q(X^N;Y^N)=H_Q(S).
\end{equation}
\end{proposition}
\begin{proof}
Suppose the plaintext $x$ was sent and the ciphertext $y$ was received, and let $s$ be their common pattern.

If $z$ has a different pattern than $x$, no key turns $z$ into $y$, so $W_N(y\mid z)=0$. If $z$ has the same pattern $s$, then $z$ uses the same number of distinct letters as $x$, say $d_s$, and by \eqref{eq:channel},
\[
W_N(y\mid z)=w_s,\qquad \text{where }w_s=\frac{(G-d_s)!}{G!}.
\]

This makes the probability of receiving $y$ easy to compute. Only plaintexts with pattern $s$ contribute, each with the same weight $w_s$, so
\[
Q_Y(y)=\sum_zQ(z)W_N(y\mid z)
=w_s\sum_{z:\,\operatorname{pat}(z)=s}Q(z)
=w_s\,Q_S(s).
\]
This is positive, because $x$ itself has pattern $s$ and $Q(x)>0$.

Therefore,
\[
\imath_Q(x;y)=\log_2\frac{W_N(y\mid x)}{Q_Y(y)}
=\log_2\frac{w_s}{w_s\,Q_S(s)}
=-\log_2Q_S(s).
\]
This holds whenever $x$ can be sent and encrypted to $y$, so $\imath_Q(X^N;Y^N)=-\log_2Q_S(S)$ almost surely. Mutual information is the average information density and since $\Pr\{X^N=x,Y^N=y\}=Q(x)W_N(y\mid x)$, we have 
\[
I_Q(X^N;Y^N)
=\sum_{x,y}Q(x)W_N(y\mid x)\log_2\frac{W_N(y\mid x)}{Q_Y(y)}
=\mathbb E\bigl[-\log_2Q_S(S)\bigr]
=H_Q(S).
\]
\end{proof}

The other side of the picture is the noise, and here a subtlety appears. The key has $H(K)\approx88.4$ bits of uncertainty for English, but not all of it affects a given ciphertext. The noise of the channel is the conditional entropy of the output given the input. Given $X^N=x$, the ciphertext is uniform over the $G!/(G-d(x))!$ strings compatible with $x$. We call
\begin{equation}\label{eq:noise}
\nu(x)=H(Y^N\mid X^N=x)=\log_2\frac{G!}{(G-d(x))!}
\end{equation}
the \emph{effective key noise} of the plaintext $x$. For example, \texttt{ATTACKATDAWN} uses only the seven letters \texttt{A, T, C, K, D, W, N}. Its ciphertext is determined by where the key sends those seven letters, so its effective key noise is $\log_2(26\cdot25\cdots20)\approx31.6$ bits; the remaining $19!$ ways of arranging the other letters are not relevant.

\begin{proposition}[Key noise that reaches the ciphertext]\label{prop:noise}
For every plaintext $x$,
\[
\nu(x)=H(K)-H(K\mid X^N=x,Y^N),
\]
so $\nu(x)\le H(K)$, with equality if and only if $x$ contains at least $G-1$ distinct letters, in which case the plaintext and ciphertext determine the key.
\end{proposition}
\begin{proof}
Fix a plaintext $x$ that uses $d=d(x)$ distinct letters. The ciphertext of $x$ depends only on where the key sends these $d$ letters. The $d$ distinct letters of $x$ can be sent to $d$ different ciphertext letters in
\[
G(G-1)\cdots(G-d+1)=\frac{G!}{(G-d)!}
\]
ways, and each choice gives a different ciphertext. Once that choice is fixed, the remaining $G-d$ letters can be arranged in $(G-d)!$ ways, and all of these keys give the same ciphertext. Since the key is uniform over all $G!$ keys and independent of the plaintext, each of the $G!/(G-d)!$ possible ciphertexts occurs with probability $(G-d)!/G!$. The ciphertext is therefore uniformly distributed over these possibilities, and
\[
H(Y^N\mid X^N=x)=\log_2\frac{G!}{(G-d)!}=\nu(x).
\]

Now suppose we also observe a compatible ciphertext $y$. By Bayes' rule and the independence of key and plaintext, a key $k$ has posterior probability
\begin{align*}
\Pr\{K=k\mid X^N=x,Y^N=y\}
&=\frac{\Pr\{K=k\mid X^N=x\}\,\Pr\{Y^N=y\mid X^N=x,K=k\}}{\Pr\{Y^N=y\mid X^N=x\}}\\
&=\frac{\Pr\{K=k\}\,\mathbf 1\{\pi_k(x)=y\}}{\Pr\{Y^N=y\mid X^N=x\}}\\
&=\frac{1/G!}{(G-d)!/G!}\,\mathbf 1\{\pi_k(x)=y\}\\
&=\frac{\mathbf 1\{\pi_k(x)=y\}}{(G-d)!}.
\end{align*}
So the key is uniform over the $(G-d)!$ keys that send $x$ to $y$, and $H(K\mid X^N=x,Y^N=y)=\log_2((G-d)!)$. This value is the same for every compatible ciphertext $y$, so averaging over $y$ gives $H(K\mid X^N=x,Y^N)=\log_2((G-d)!)$.

Subtracting from $H(K)=\log_2(G!)$,
\[
H(K)-H(K\mid X^N=x,Y^N)=\log_2(G!)-\log_2((G-d)!)=\nu(x).
\]
\end{proof}

Proposition~\ref{prop:noise} splits the key's uncertainty into two parts:
\[
H(K)=\underbrace{\nu(x)}_{\text{reaches the ciphertext}}+\underbrace{\log_2((G-d)!)}_{\text{stays hidden}}.
\]
The first part concerns the letters that occur in the plaintext. Where the key sends them is visible in the ciphertext, so this part acts as noise on the channel. The second part concerns the letters that never occur. Nothing in the ciphertext depends on them, so they remain unknown even to someone who sees both the plaintext and the ciphertext. 

The two parts trade off as the plaintext uses more of the alphabet. Writing out the factorials,
\[
\nu(x)=\log_2\bigl(G(G-1)\cdots(G-d+1)\bigr)=\sum_{j=0}^{d-1}\log_2(G-j),
\]
so each new letter moves some uncertainty from the hidden part to the noise. The first letter of the plaintext can be sent to any of $G$ ciphertext letters, so it adds $\log_2G\approx4.70$ bits of noise. The second distinct letter can go to any of the remaining $G-1$, adding $\log_2 25\approx4.64$ bits, and in general the $(j+1)$st distinct letter adds $\log_2(G-j)$ bits. The increments shrink because each new letter has fewer places left to go. The last one adds nothing, since once $G-1$ letters are placed, the final letter is determined. 

\subsection{Capacity and the Channel Coding Theorem}\label{sec:capacity}
We now know what the substitution channel transmits (the pattern of repeated letters) and how much noise the key adds. The natural next question in communication theory is: how many different messages can the channel carry without the receiver confusing them? 

The channel coding theorem answers this question when the noise is redrawn independently for every transmission. Suppose, then, that every new message of length $N$ is encrypted with an independently chosen key, as when two parties agree to change the key for every message. The channel transitions across successive blocks are then conditionally independent given the inputs, so the encryption system is a discrete memoryless channel whose input is a whole plaintext of $N$ letters and whose output is a whole ciphertext. Its \emph{channel capacity}, in bits per ciphertext, is
\[
C_N=\max_QI_Q(X^N;Y^N),
\]
where the maximum ranges over all input laws $Q$ on $\mathcal X^N$. 

To bound $C_N$ from below, we need one specific input law, and the simplest is the uniform law on all sequences of $N$ letters. Under this law the information density takes a very simple form.

\begin{lemma}[Information density under the uniform law]\label{lem:uniform}
Let $Q$ be the uniform law on all $G^N$ sequences of $N$ letters. Then every compatible pair of plaintext and ciphertext $(x,y)$ satisfies
\begin{equation}\label{eq:uniform-density}
\imath_Q(x;y)=N\log_2G-\nu(x).
\end{equation}
\end{lemma}
\begin{proof}
By Proposition~\ref{prop:information}, $\imath_Q(x;y)=-\log_2Q_S(s)$, where $s$ is the pattern of $x$. A pattern $s$ with $d$ distinct letters is shared by exactly $G!/(G-d)!$ plaintexts. Under the uniform law, $Q_S(s)=\frac{G!/(G-d)!}{G^N}=2^{\nu(x)-N\log_2G}$, which gives \eqref{eq:uniform-density}.
\end{proof}

This gives a lower bound on channel capacity.

\begin{proposition}[Capacity lower bound]\label{prop:capacity}
\begin{equation}\label{eq:capacity}
C_N\ge N\log_2G-H(K).
\end{equation}
\end{proposition}
\begin{proof}
Take $Q$ uniform on $\mathcal X^N$. By Lemma~\ref{lem:uniform} and Proposition~\ref{prop:noise}, every compatible pair of plaintext and ciphertext has $\imath_Q(x;y)=N\log_2G-\nu(x)\ge N\log_2G-H(K)$. Averaging gives $I_Q(X^N;Y^N)\ge N\log_2G-H(K)$, and $C_N$ is at least this value.
\end{proof}

To state the coding theorem, we need the notion of a code. A \emph{code} over $m$ channel uses is a list of $M_m$ messages, agreed on in advance and known to everyone, including the attacker, in which each message has $mN$ letters divided into $m$ plaintexts of $N$ letters. To send a message, each of its plaintexts is encrypted with an independent fresh key. The attacker knows the list and observes all $m$ ciphertexts, but none of the keys, and must decide which plaintexts were sent. The \emph{rate} of the code, $(\log_2M_m)/m$, is the number of bits of plaintext carried per ciphertext.

The channel coding theorem \cite{cover2006} states that for every rate $0\le r<C_N$ there are codes whose rates tend to $r$ and whose average decoding error tends to zero as $m\to\infty$, while rates above $C_N$ cannot have vanishing error.

The typical-set approximation suggests a target rate of $NR$ bits per ciphertext. This rate can be achieved by suitably chosen codes over many blocks when $NR<C_N$; the theorem does not guarantee recovery for a prescribed list of English texts. By Proposition~\ref{prop:capacity}, this holds as soon as $NR<N\log_2G-H(K)$, that is, since $D=\log_2G-R$,
\[
ND>H(K)\quad\Longrightarrow\quad NR<C_N.
\]

This recovers the classical scale $H(K)/D$ \cite{hellman1977} as a sufficient capacity comparison.

In other words, if $ND>H(K)$, one can choose a list of approximately $2^{mNR}$ messages, each consisting of $m$ length-$N$ plaintext blocks, whose average decoding error tends to zero as $m$ grows. Each of the $m$ blocks is encrypted with its own independent key. The attacker knows the chosen list and jointly decodes the $m$ ciphertexts without knowing the keys. 

The channel coding theorem recovers the classical expression as a sufficient condition for reliable communication using suitably chosen codebooks. We now examine how well this expression describes unique decipherment in a fixed collection of English passages, and compare it with predictions from N-gram language models.

\section{An Experiment with Six English Novels}
\label{sec:experiment}

The classical formula gives approximately 28 letters. We now
ask how well this predicts unique decipherment when the
plaintext is a passage from a known collection of books.

\subsection{The Experiment}

Imagine that the sender and the attacker own the same six
novels: \emph{Alice's Adventures in Wonderland},
\emph{Pride and Prejudice}, \emph{Frankenstein},
\emph{Moby Dick}, \emph{The Adventures of Sherlock Holmes},
and \emph{Adventures of Huckleberry Finn}.
The sender chooses a passage and encrypts it using a simple
substitution cipher. The attacker knows that the plaintext
comes from these books, but does not know the passage or
the key.

We remove spaces, punctuation, and other nonletter characters
and convert all letters to uppercase. For each length
$N=10,\ldots,24$, we collect every passage of $N$ consecutive
letters within a book and remove duplicate strings.
Let $\mathcal C_N$ be the resulting set and
$M_N=|\mathcal C_N|$. The sender chooses uniformly from
$\mathcal C_N$.

A substitution preserves which positions contain the same
letter. For example, \texttt{EYE} and \texttt{DID} both have
pattern $(1,2,1)$ and can produce the same ciphertext under
different keys. A passage is therefore uniquely identifiable
exactly when no other passage in $\mathcal C_N$ has its
pattern. We check every passage and calculate
\[
u_N=
\frac{\text{number of passages with a unique pattern}}{M_N}.
\]
This gives the \textit{exact} probability of unique plaintext recovery
within the six-book collection, without relying on a particular
decoding algorithm.

\subsection{Predictions from Models of English}

We compare the observed probabilities with four models
based on public English letter-frequency tables
\cite{burkardt}. An \emph{N-gram} is a sequence of $N$
consecutive letters: monograms are single letters,
bigrams are pairs, trigrams are triples, and quadgrams
are groups of four letters.

Let $c(w)$ be the published count of a letter sequence $w$.
We convert these counts into probabilities:
\begin{align*}
p_a&=\frac{c(a)}{\sum_b c(b)},&
T_{ab}&=\frac{c(ab)}{\sum_c c(ac)},\\
A_{ab,c}&=\frac{c(abc)}{\sum_d c(abd)},&
B_{abc,d}&=\frac{c(abcd)}{\sum_e c(abce)}.
\end{align*}
Here $p_a$ is the probability of letter $a$, while
$T_{ab}$, $A_{ab,c}$, and $B_{abc,d}$ give the probability
of the next letter given the preceding one, two, or
three letters.

For a string $x=(x_1,\ldots,x_N)$, the four models are
\begin{align*}
Q_0(x)&=\prod_{t=1}^N p_{x_t},\\
Q_1(x)&=p_{x_1}\prod_{t=2}^N T_{x_{t-1},x_t},\\
Q_2(x)&=p_{x_1}T_{x_1x_2}
          \prod_{t=3}^N A_{x_{t-2}x_{t-1},x_t},\\
Q_3(x)&=p_{x_1}T_{x_1x_2}A_{x_1x_2,x_3}
          \prod_{t=4}^N B_{x_{t-3}x_{t-2}x_{t-1},x_t}.
\end{align*}
Thus $Q_0$ generates letters independently, while the
other models use progressively longer preceding sequences.
Unlisted sequences have count zero. If a denominator is
zero, we use the longest shorter suffix of the preceding
sequence for which the denominator is positive, falling
back to single-letter probabilities if necessary.

For each model and length $N$, we independently generate
$M_N$ strings and calculate the fraction with no different
string sharing their repetition pattern. Identical copies
count separately toward this fraction but do not make
a plaintext ambiguous. We average the results over five
independent runs. We also include a model with independent,
uniformly distributed letters, whose recovery probability
can be calculated directly.

These comparisons hold the number of candidates fixed
while changing how their text is generated. They test
whether short letter sequences explain the ambiguity
found in real passages. The code and data are available on
\href{https://github.com/fangyuanlin2002/substitution-unicity-distance}{GitHub}.

\subsection{Results}

\begin{figure}[htbp]
\centering
\includegraphics[width=\textwidth]{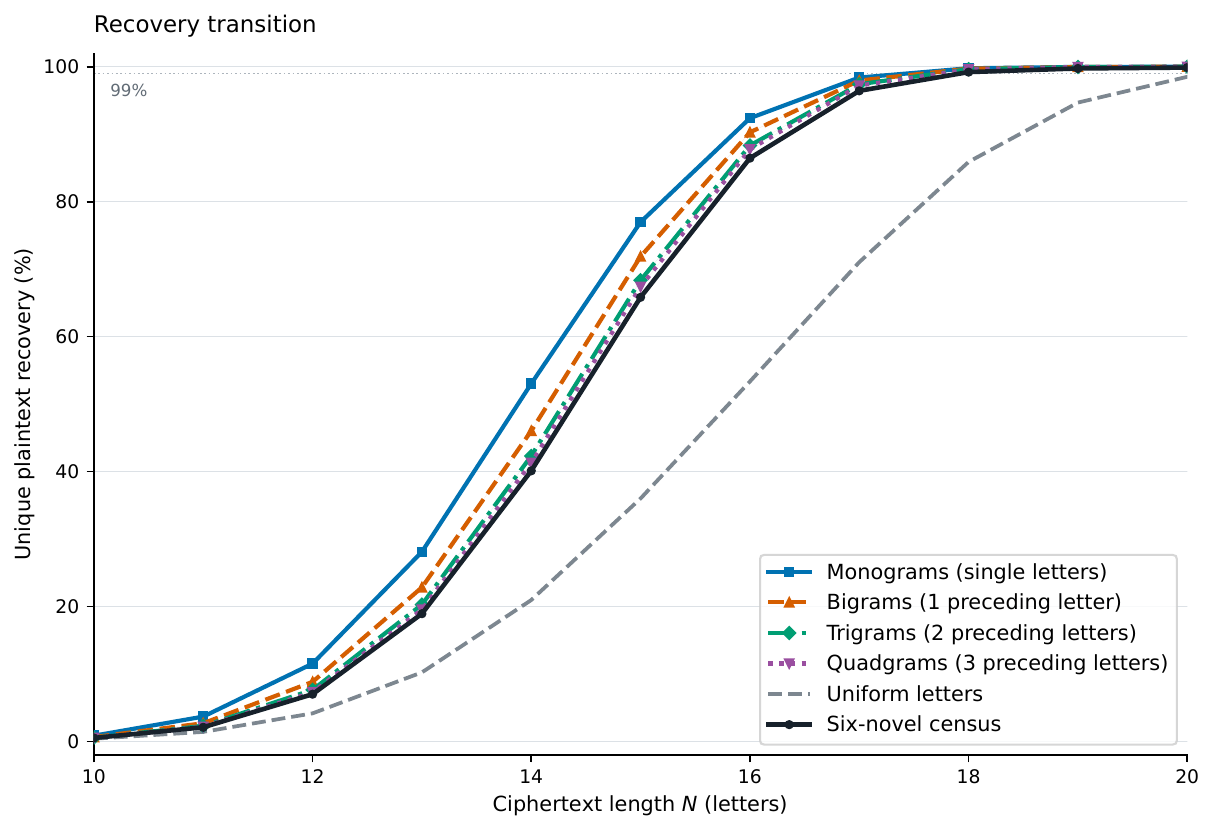}
\caption{Probability of unique plaintext recovery.
The six-novel curve is calculated exactly; the English-model
curves average five simulations. The uniform-letter curve
is calculated directly. The classical estimate of about
28 letters lies beyond the displayed range.}
\label{fig:recovery}
\end{figure}

Figure~\ref{fig:recovery} compares the exact recovery
probabilities for the six novels with the model predictions. At 18 letters, $99.2075\%$ of the $2{,}772{,}123$ distinct
passages are uniquely identifiable. All four N-gram models
also first exceed $99\%$ recovery at 18 letters among the
tested lengths. The uniform-letter model reaches this level
at 21 letters.

Using longer letter sequences improves the predictions.
Across the tested lengths, the largest differences from
the observed recovery probability are 12.91, 6.04, 2.52,
and 1.56 percentage points for monograms, bigrams, trigrams,
and quadgrams, respectively. Simulation uncertainty is small
compared with these differences.

\begin{figure}[htbp]
\centering
\includegraphics[width=\textwidth]{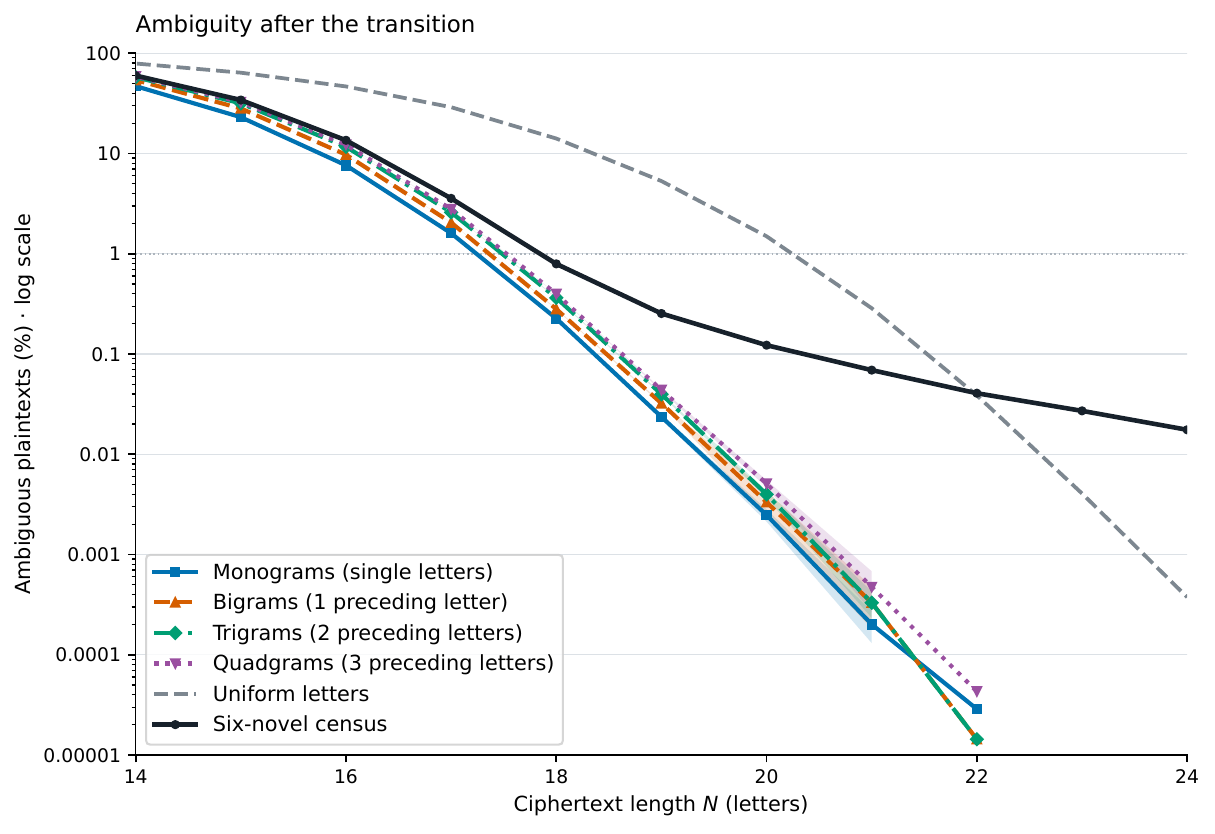}
\caption{Probability that another plaintext remains compatible
with the ciphertext. The logarithmic scale makes small
probabilities easier to compare. Shading shows $\pm2$
estimated standard errors for the simulated curves.}
\label{fig:ambiguity}
\end{figure}

Close agreement in recovery probability can hide differences
in the ambiguity that remains (Figure~\ref{fig:ambiguity}). At 18 letters, $0.7925\%$ of the real passages still have
another compatible reading, compared with $0.3972\%$ under
the quadgram model. Thus even the best-fitting model
underestimates the remaining ambiguity by about a factor
of two. Real passages can overlap and share phrases,
features that independently generated strings do not preserve.

The collection exceeds $99\%$ unique plaintext recovery at
18 letters, below the classical estimate of about 28.
The classical calculation counts the number of spurious keys, whereas
our experiment counts distinct compatible plaintexts. Hellman warns that counting keys rather than messages can
make a cipher look more secure than it is
\cite{hellman1977}. An 18-letter passage leaves at least eight letters unobserved, so even a uniquely recovered plaintext admits at least
$8!=40{,}320$ compatible keys.

The two lengths describe different quantities. $H(K)/D$ is the approximate length at which the expected number of spurious keys falls to one. Our channel interpretation also shows that
$N>H(K)/D$ is sufficient for reliable communication at
rate $NR$ bits per block using suitably chosen codebooks
over independently encrypted blocks. By contrast, 18 letters is the first tested length at which more than $99\%$ of the distinct passages in our six-novel collection are uniquely determined by the ciphertext.
It means that exactly one candidate plaintext
is compatible with the observed ciphertext, whether or not
a particular solver finds it.

Changing the numerical entropy rate alone cannot make
the classical expression equal this empirical length.
For $0<D\le\log_2 26\approx4.70$,
$H(K)/D\ge88.38/\log_2 26\approx18.8$;
setting $H(K)/D=18$ would require a negative entropy rate
$R\approx-0.21$. Even substituting the collection's
finite-codebook rate
$R_{18}=\log_2(2{,}772{,}123)/18\approx1.19$ bits per letter
gives $H(K)/(\log_2 26-R_{18})\approx25$.

The N-gram models closely reproduce the exact recovery
probabilities for the six-novel collection, giving a better
and better fit as the N-gram order increases (from monograms to quadgrams). The increasingly close agreement with models built from \textit{public} English N-gram frequencies suggests that the observed transition at 18-letter may reflect broader regularities of English, rather than peculiarities of these six novels. The sufficient coding condition in Section~\ref{sec:capacity}
remains valid, but does not locate this collection's
plaintext-recovery threshold.

\section{Conclusion}
We started by treating encryption as transmission through a noisy channel. Applying the channel coding theorem led back to the classical expression $H(K)/D$. We then asked whether that expression describes the recovery of passages from six known novels. It does not: more than $99\%$ of the passages are uniquely identifiable at 18 letters, well before the classical estimate of about 28. Models based on short English letter sequences predict the observed probabilities much more closely, though they still underestimate the ambiguity that remains.

Unique decipherment depends on whether another possible plaintext fits the same ciphertext. In our experiment, the language’s entropy rate alone does not predict how often this happens. The channel interpretation explains the classical formula, while N-gram models give a more accurate account of the observed recovery probabilities.

\section*{Acknowledgments}
The author thanks Professor Steven N. Evans for his guidance and support, Professor Andrew Gelman for helpful discussions, and the reviewers and editor of the PUMP Journal of Undergraduate Research for their patience and thoughtful suggestions.

\end{document}